# Observational Evidences of the Cosmological Deceleration of Time

© I.N. Taganov[1,2]

1. Russian Geographical Society, Saint Petersburg, Russia
2. taganov.igor@mail.ru

**Abstract.** Constancy of the speed of light together with the Hubble law lead in a doctrine of expanding universe to a conclusion that universe evolution is not only an expansion of space but also a deceleration of the course of physical time (Taganov, 2003). Duration of any physical process measured by decelerating physical time $(\tau)$ is always longer, than corresponding duration, measured by a scale of invariable uniform Newtonian time $(t)$: $\tau = t + Ht^2/2$. All the processes uniform or moderately decelerating in physical time appear accelerating when interpreted in terms of uniform invariable Newtonian time. The article describes several observational evidences of a phenomenon of the cosmological deceleration of time: the redshift survey of distant supernovae demonstrates apparent acceleration of the universe expansion; accelerations in the Earth-Moon system reveal reliable differences if estimated with the use of ancient and contemporary astronomical data. Deceleration of the physical time reveals itself in isotopic chronology by systematic divergence of samples Newtonian ages, estimated with the use of isotopes with different rates of decay.

## Introduction

A doctrine of "expanding" space of the universe $\Delta l(\tau) = a(\tau)\Delta l_0$ with monotone increasing scale-factor $a(\tau)$ together with a condition of the constancy of the speed of light: $\Delta l = c\Delta\tau; \Delta l_0 = c\Delta\tau_0; c = const$ lead to the relation:

$$\Delta\tau(\tau) = a(\tau)\Delta\tau_0 \qquad (1)$$

This relation suggests that time also "expands" along with the space in the course of universe evolution. While a term "space expansion" is a common cosmological term today, the somewhat clumsy term "time expansion" is better to replace with the more accurate term "deceleration of the course of time". The course of time is defined as a value $\Delta\tau^{-1}$, converse to the chosen time standard $\Delta\tau$. Increasing time standard corresponds to decreasing course of time and thus to deceleration of the course of time. The course of time concept was probably first formulated by Einstein and Minkowski in their pioneering works in the relativity theory. The term "course of time" was later favored by J. Synge [4] and N.A. Kozyrev [7].

The decelerating, expanding time in (1) cannot be Newtonian time commonly used by natural science as invariable homogeneous continuum. The time in (1) displaying a deceleration in the process of universe evolution is referred to in this article as "physical". The term "physical time" is justified by analogy with the term "physical vacuum" used by quantum physics instead of the old classical concept of "emptiness" as an abstract three-dimensional mathematical continuum. Quantum physics defines vacuum state by fluctuations of interacting quantum fields. These fluctuations correspond to zero-oscillations in quantum mechanics and govern multiple transformations of virtual micro-particles resulting, in particular, in physical vacuum polarization. The fluctuation spectrum change and vacuum polarization in volumes with electro-conducting boundaries are made evident by Casimir macroscopic forces, independent of masses, charges or any other coupling factors. Since modern physics accepts a conception of non-stationary space-time, a principle of the constancy of the speed of light and quantum postulates, the decelerating time rightfully can be referred to as physical. Similar to the physical vacuum theory, a conception of the cosmological deceleration of the course of time is substantiated with relativistic and quantum ideology. Physical time is henceforth symbolized by $\tau$, Newtonian time by $t$, with $a' = da/d\tau$ and $\dot{a} = da/dt$.

The relation defining changing scale of physical time in a non-stationary universe can be derived from Hubble law considered as a purely empirical relation. According to Hubble law photon wavelength varies as: $\lambda = (1 + Ht)\lambda_0$ and a relation between photon wavelength and the period is: $\lambda_0 = c\Delta t$. For big cosmological distances and, respectively, the big time intervals, relation defining increasing photon wavelength can be used as the differential equation: $dl = (1 + Ht)dl_0 = (1 + Ht) \cdot cdt$. For finite distances described as: $l = c\tau$, integration of this equation gives: $l = c\tau = c\int_0^t (1 + H\xi)d\xi$ and respectively:

$$\tau = \int_0^t (1+H\xi)d\xi = t + H/2\, t^2 \tag{2}$$

Geometrical interpretation of physical time is ascertained by possibility to interpret a non-stationary metric with interval: $ds^2 = -c^2 d\tau^2 + a^2 dr^2$ as a result of conformal transformation of Minkowski metric (see e.g. [4]). The transformation $t = \int \exp[-\psi(\tau)/2]d\tau$ with $\psi(\tau) = 2\ln a$ is matched by: $d\tau^2 = \exp[\psi(\tau)]dt^2 = a^2 dt^2$, allowing to represent the non-stationary interval as: $ds^2 = a^2(-c^2 dt^2 + dr^2)$. With this interval definition, time-dependent scale factor does not influence the velocity at radial world lines, as their intervals when $ds = 0$ correspond to Minkowski metric.

Relation (2) demonstrates that duration of any physical process measured by the decelerating physical time is always longer, than corresponding duration, measured by a scale of invariable uniform Newtonian time. For example, for the theoretical estimation of Hubble constant (Taganov, 2008 [6]): $H = 9\hbar G/16 c^2 r_e^3 = 1.970 \cdot 10^{-18}$ s$^{-1}$ (61.6 km/s/Mpc) one can estimate Newtonian universe age as: $t_p = H^{-1} = 5.081 \cdot 10^{17} s = 16.131$ Gyr. In accordance with Eq. 2 the corresponding physical age of the universe is: $\tau_p = 3/2H = 7.614 \cdot 10^{17} s = 24.1$ Gyr.

Observational data discussed in this article suggest that all physical processes in a non-stationary universe evolve in the cosmologically decelerating physical time and just this time should be used in mathematical models of the large-scale structure of Metagalaxy.

## 1. Illusion of accelerating universe expansion and the "dark energy"

In the past decade, astrophysical journals were engaged in a lively discussion of the observational evidence of acceleration in the universe expansion, found by two independent research teams (see e.g. [2, 3]). This phenomenon was discovered using estimates of cosmological parameters $\Omega_m, \Omega_\Lambda$ in the Standard model of classical cosmology, derived from the analysis of supernovae redshift survey. From observational data analysis authors conclude that the deceleration parameter is reliably negative ($q_{pe} < 0$) and, consequently, the universe expansion accelerates in our epoch. This interpretation of supernova redshifts complements with another impressive observations suggesting accelerating expansion of the universe in our epoch. The HKP (Hubble Key Program) for Cepheid studies in galaxies at a distance up to 20 Mpc ($z < 0.1$) estimated the Hubble parameter as: $H_e = (72 \pm 8)$ km/s/Mpc. On the other hand, redshift study of supernovae further in the past ($z = 0.1 \div 0.6$) yield smaller value of Hubble parameter: $H_e = (65 \pm 7)$ km/s/Mpc [3]. This difference in the values of observed Hubble parameter provides another argument for an assumption of the accelerating universe expansion in recent times.

Assuming the existence of enigmatic "dark energy" is the most popular way in classical cosmology to explain recent observations that the universe appears to be expanding at an accelerating rate. In classical cosmology dark energy is a hypothetical form of energy that permeates all the space and tends to increase the rate of the universe expansion. In the Standard model of classical cosmology, dark energy currently accounts for almost three-quarters of the total mass-energy of the universe. Two proposed forms of dark energy are:
- the cosmological constant, representing a constant energy density filling space homogeneously, and
- scalar fields such as "quintessence" or "moduli", dynamic quantities whose energy density can vary in time and space.

In fact, contributions from scalar fields that are constant in space are usually also included in the cosmological constant. Scalar fields, which do change in space, are hard to distinguish from a cosmological constant, because the change may be extremely slow.

Since the dark energy was included in the Standard cosmological model exclusively to explain the accelerating universe expansion, this ghostly substance will disappear at once after demonstration of the illusory acceleration essence. In the conception of decelerating physical time the observed accelerating universe expansion has simple explanation. With the help of formula $\dot{a} = aa'$, one can obtain a relation between cosmological deceleration parameter in physical time $q_\tau$ and the deceleration parameter $q_t = -a\ddot{a}/\dot{a}^2$ in Newtonian time:

$$q_t = q_\tau - 1 \qquad (3)$$

It follows from this relation that moderately decelerating ($q_\tau < 1$) universe expansion in physical time appears as accelerating expansion ($q_t < 0$) in Newtonian time. Uniform expansion in Newtonian time ($q_t = 0$) corresponds to decelerating expansion in physical time ($q_\tau = 1$). These coordinate effects are similar to kinematics changes in the transitions between inertial and non-inertial frames of reference.

To clarify the essence of the illusory accelerating universe expansion, one should bear in mind that Hubble law is exact only for Newtonian time [5, 6]. On the other hand, estimations of the Hubble parameter from astrophysical observations of the processes evolving in physical time actually make use of the approximate relation: $z = Ht \simeq H^*\tau$. The relation between exact value of the Hubble parameter $H$ and its approximate estimate $H^*$ can be derived from Eq. 2: $z = Ht \simeq H^*\tau = H^*(t + Ht^2/2)$. Substituting in this relation the Hubble law in the form: $t = z/H$ one can get the formula:

$$H^* = 2H/(2+z) \qquad (4)$$

It follows from this relation that the approximate value of the Hubble parameter $H^*$ diminishes with the increase of cosmic object redshifts used to evaluate this parameter. Eq. 4 can be compared with observed redshifts in supernovae spectra. In Fig. 1 filled circles correspond to estimated Hubble parameters $H_e^*$ in several redshift ranges, characterized by the mean redshifts $<z> = (z_{max} + z_{min})/2$. Descending dotted curve in Fig. 1 corresponds to Eq. 4 with theoretical value of Hubble constant [6, 11]: $H = 9\hbar G/16c^2 r_e^3 = 1.970 \cdot 10^{-18}$ s$^{-1}$ (61.6 km/s/Mpc) and coincides with the observational data with relative standard deviation below 7 %. Fig. 1 demonstrates how an illusion of the accelerating universe expansion emerges, when not exact, but approximate values of Hubble parameter evaluated from the set of increasing redshifts: the evaluations of Hubble constant for increasing redshifts (apparent journey in the past) give decreasing values (declining rate of universe expansion).

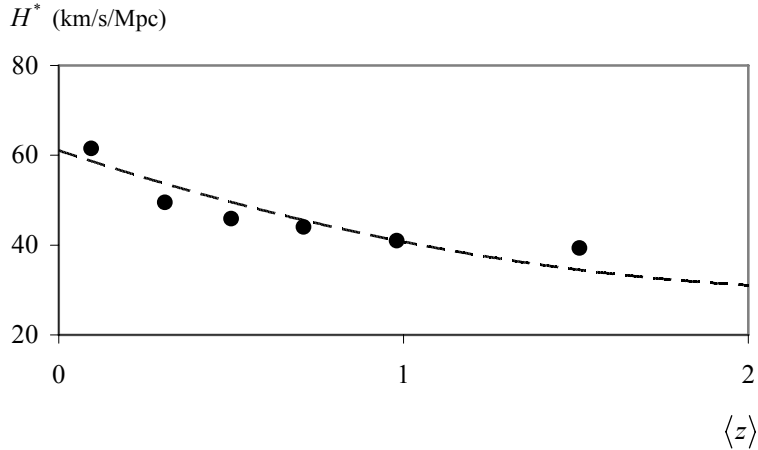

*Fig.1 The observational estimations of Hubble parameter (filled circles) compared with Eq. 4 (dotted curve).*

Authors of the article [3] present the presumable kinematics of the universe expansion based on analysis of the supernova redshift survey. This analysis uses the evolution equation in the form of three terms in Taylor series:

$$a(t) = a_0[1 + H(t - t_0) - \tfrac{1}{2} q_0 H^2 (t - t_0)^2] \qquad (5)$$

and the deceleration parameter-redshift dependence represented as the two-parameter equation:

$$q(z) = q_0 + z\, dq/dz \qquad (6)$$

The value of $dq/dz$ is defined at $z = 0$. The "transition redshift" $z_t$ used as characteristic of the epoch when decelerating expansion transforms to accelerating one is:

$$z_t = -q_0/(dq/dz) \tag{7}$$

The kinematics parameter estimations from observation depend on the adopted confidence limit as listed in Table 1 (from Fig. 5 in [3]).

*Table 1. Kinematics parameters of the "accelerating" universe expansion*

| Confidence limit | $q_0$ | $dq/dz$ | $z_t$ |
|---|---|---|---|
| 99 % | $-0.85^{+0.5}_{-0.5}$ | $2^{+2}_{-1.9}$ | $4.3^{+10}_{-4.2}$ |
| 95 % | $-0.87^{+0.5}_{-0.5}$ | $2^{+1.5}_{-1.5}$ | $1^{+1.7}_{-0.9}$ |
| 68 % | $-0.82^{+0.3}_{-0.3}$ | $1.9^{+0.9}_{-0.9}$ | $1.2^{+1.9}_{-1}$ |

According to the authors opinion [3], their estimates of kinematics parameters indicate the accelerating universe expansion ($q < 0$) in recent times and decelerating expansion ($q > 0$) in more early epochs at $z > 1$.

The illusiveness of this kinematics can be demonstrated by the analysis of the *uniform* universe expansion with *zero acceleration*. In accordance with Hubble law for the uniform expansion it holds: $a(t) = a_0[1 + H(t - t_0)]$. If we acquire $t_0 = 0$, then for past epochs ($t < 0$) this relation is: $a(t) = a_0[1 - Ht]$. If we take into account the difference between Newtonian and physical time then the approximate Hubble parameter is defined by Eq. 4 and the evolution equation for uniform expansion under discussion will look like:

$$a(t) = a_0(1 - H^*t) = a_0[1 - 2Ht/(2 + z)]) \tag{8}$$

However, this equation describes now a *non-uniform expansion* with *non-zero acceleration*. To estimate kinematics parameters of the apparent acceleration, one should compare Eq. 8 with Eq. 5 for the past epoch:

$$a(t) = a_0(1 - Ht - \tfrac{1}{2}q^* H^2 t^2) \tag{9}$$

Equating (8) and (9), taking into account that from $z = H^*t$ it follows: $Ht = z(2 + z)/2$, one can obtain:

$$q^*(z) = -4/(2 + z)^2 \tag{10}$$

From this relation at $z = 0$ one can obtain the following values for the kinematics parameters:

$$q_0^* = -1 \quad dq^*/dz = 1 \quad z_t^* = 1 \tag{11}$$

Comparing these values with the observational estimates presented in Table 1 one can be convinced that they agree with accuracy of one standard deviation. This analysis demonstrates that the uniform universe expansion in the idealized Newtonian time appears as a non-uniform expansion with variable acceleration when kinematics parameters are estimated from the observational data for astrophysical processes evolving in true decelerating physical time.

The quantum cosmological model [5, 6] describes not illusory but true decelerating universe expansion in physical time, corresponding to the constant deceleration parameter: $q_\tau = +1$. In our epoch a deceleration of the universe expansion has an order of $-H^2$ s$^{-2}$.

The illusion of accelerating expansion of the universe will emerge always when for interpretation of the astrophysical processes evolving in decelerating physical time the cosmological models with uniform invariable Newtonian time are used.

## 2. Strange accelerations of the Earth and the Moon

Secular acceleration of the Moon was discovered in 1695 by Edmond Halley when he was comparing own observations of the Moon with astronomical data about eclipses of the Moon in ancient times. Estimation of the Moon acceleration obtained in the 18$^{th}$ century by several astronomers confirmed Halley' findings, and Damoisie was the first to calculate that secular longitude increase of the Moon in orbit is not too big being about $10''$ (seconds of arc) per century. It means that if in our epoch the Julian year, containing 365.25 ephemeris days ($3.15 \cdot 10^7$ s), consists of 13.37 lunar sidereal months (27.32 days $\simeq 2.36 \cdot 10^6$ s $\simeq 1.3'' \cdot 10^6$), thus an additional lunar month in a year will appear only in next 15 million years.

In 1754 Immanuel Kant proposed a hypothesis about possible deceleration of the Earth rotation and apparent acceleration of the Moon motion due to tides caused by the Sun and Moon on the Earth. Laplace, however, gave an outright denial of this Kant' idea because it led to conclusion about secular accelerations of the Earth and planets that then had not been yet discovered by astronomers.

In 1770 the Academy of Sciences in Paris launched a competition for the best explanation of the Moon acceleration. The theory created by Laplace in 1787 won recognition. Laplace supposed that gravitational disturbances from the planets lead to gradual decrease of the Earth orbit eccentricity that has an influence on the perturbation of the Moon orbit by the Sun and gradually increases the orbital velocity of the Moon. Laplace demonstrated amazingly high accuracy of his theory having evaluated the Moon acceleration as $10''.18$ per century. It coincided almost exactly with the observational data. Having used Laplace' gravitational concept several astronomers evaluated their own theoretical estimations of the Moon acceleration per century - Damoisie: $10''.7$; Plana: $10''.6$; Ponteculan: $12''.24$; Hansen: $12''.18$.

Laplace' gravitational theory was considered the most complete description of the Moon acceleration up to 1853. In this year, John Adams, known for his prediction of the Neptune existence made by him independently of Le Verie, found errors in the calculations made by Laplace and his successors. Having corrected Laplace' mistakes he evaluated the gravitational acceleration of the Moon as $5''.7$ per century proving that Laplace' gravitational theory could describe only about half of the observed Moon acceleration. Upon correction of Laplace' mistakes found by Adams, Charle Delonet estimated the theoretical value of the Moon gravitational acceleration as $6''.18$. After rather a long discussion Plana and Hansen also acknowledged Adam' calculations and agreed that Laplace' gravitational theory could only explain around $6''.1$ of the observed Moon acceleration. Hitherto this value is accepted as the theoretical estimate of the part of the Moon acceleration caused by gravitational perturbations in the Solar system. Adams takes credit not only for the revision of Laplace' estimates; he proved that the gravitational concept en principle cannot explain all the Moon acceleration. John Adams supposed that the Moon acceleration depends not only on gravitational forces but also on some other factors, prophesying: "This fact can lead us to an important physical discovery".

In the 1870s to develop an accurate theory of the Moon motion including observed acceleration was used an old Kant' theory about the influence of earth tides on the Moon and Earth motion. The Earth-Moon gravitational force creates a tidal wave that causes a local rise in water basins of the Earth and mechanical stresses in the earth crust. This tidal wave follows the Moon lagging behind it a little bit because of its energy dissipation. The tidal wave is moving against the Earth rotation and because of friction decelerates the rotation. However, the tidal wave influences also the Moon motion. On the rotating Earth the tidal wave creates the momentum that decelerates the Moon motion and increases its orbital semi-axis. Therefore, the tidal waves decelerate rotation both the Moon and the Earth. Thus, on the one hand, tidal waves cause true deceleration of the Moon rotation. On the other hand, deceleration of the Earth rotation leads to an apparent acceleration of the Moon motion around the Earth. If the Earth rotation deceleration by tidal waves exceeds the true deceleration of the Moon, we will observe an apparent acceleration of the Moon motion in orbit.

Laplace' gravitational theory and Kant' tidal mechanism laid the foundation to the so-named Hill-Brown theory of the Moon motion developed in 1878 – 1909 (see e.g. the modern version of this theory in [8]). The modern version of Hill-Brown theory explains $6''.1$ of the Moon acceleration by Laplace' gravitational concept and about $5''$ by Kant' tidal mechanism. Because of incompleteness of the tide geophysical theory it has not been possible yet to evaluate exactly the tidal momentum and dissipation of tidal wave energy and therefore Hill-Brown theory uses semi-empirical equations.

Astronomy determines the longitude of the planet in orbit in seconds of arc using the following equation:

$$L = L_0 + nt_c + \delta L \qquad (12)$$

In this equation $t_c$ is Newtonian (ephemeral) time measured in Julian centuries (36,525 days) starting from the "fundamental epoch": mean Greenwich time, midday of January 0, 1900 (hereinafter the "c" index stands for the time intervals measured in Julian centuries); $L_0$ is the planet longitude at the zero time-reading; $n$ is the mean planet motion in seconds of arc per century corresponding to Kepler and Newton laws, and $\delta L$ stands for various observed deviations from the mean motion. Astronomy uses the terms "solar motion" and "solar acceleration" having in mind characteristics of the observed Sun motion across the sky. Therefore, for example, "solar acceleration" $\dot{\sigma}_0$ is proportional to angular acceleration of the Earth rotation (see Eq. 14).

The theory of the Moon tidal acceleration employs Kepler and Newton laws, the principle of angular momentum conservation in the Earth-Moon system and the law of energy conservation. For angular acceleration of the Moon $\dot{n}$ and rotational acceleration of the Earth $\dot{\omega}$ the following equations are used (see e.g. [1]):

$$\dot{n} = -3N/mb^2 \qquad \dot{\omega} = -(N + N_0)/I \tag{13}$$

In these equations $m$ stands for the Moon mass, $b$ is the mean semi-axis of the lunar orbit, $N$ and $N_0$ are tidal force momentums affecting the Earth from the Moon and the Sun respectively; $I \simeq MR^2/3$ is the Earth momentum of inertia. Accelerations of the Moon $\dot{\sigma}$ and the Sun $\dot{\sigma}_0$ observed from the Earth are determined by the formulae:

$$\dot{\sigma} = \dot{n} - n\dot{\omega}/\omega \qquad \dot{\sigma}_0 = \dot{n}_0 - n_0\dot{\omega}/\omega \simeq -n_0\dot{\omega}/\omega \tag{14}$$

Using Eqs. 13, 14 one can derive the relation:

$$\dot{\sigma}/\dot{\sigma}_0 = n/n_0 [1 - \alpha \cdot 3I\omega/mb^2 n] = n/n_0 (1 - 0.62 \cdot \alpha) \qquad \alpha = N/(N + N_0) \tag{15}$$

Here $n/n_0 = 13.37$ is a relation of the average motions of the Moon and the Sun, i.e. the number of sidereal lunar months in a year; $\omega/n = 27.32$ is the relation of average motions of the Earth and the Moon, i.e. the number of days in the sidereal lunar month. Parameter $\alpha = N/(N + N_0)$ defines a share of the lunar tides in total tide influence in the Earth-Moon system. This parameter changes from 0 (tides absence) to 1 (only lunar tides).

At the same time with development of the tidal theory of the Moon acceleration, several astronomers started labor-consuming work to estimate the value of observed acceleration of the Moon motion. The specific feature of this work was the use of two groups of astronomical data: what is known as "antique" astronomical data from between the beginning of the first millennium BC and the first centuries AD and "modern" observations systematized approximately for the period of 1700 - 1950. The difference between antique and modern epochs of observations is considered to be about 2000 years. In Table 2 listed estimations of solar and lunar tidal accelerations (minus lunar gravitational acceleration $6''.1$ per century), based on the analysis of the modern and antique astronomical data.

*Table 2. Estimations of accelerations in the Earth-Moon system*

| | | Fotheringham, J [2] | De Sitter, W [1] | Sir H. Spencer [3] | Newton, R [4] |
|---|---|---|---|---|---|
| Moon acceleration $\dot{\sigma}$ ($c^{-2}$) | Antique observations | $-4''.7$ | $-5''.22 \pm 0''.30$ | | $-9''.36 \pm 3''.16$ $(-14''.70 \pm 2''.15)$ |
| | Modern observations | | | $(-11''.22 \pm 1'')$ | $(-10'' \pm 1''.50)$ |
| Sun acceleration $\dot{\sigma}_0$ ($c^{-2}$) | Antique observations | $1''.5$ | $1''.8 \pm 0''.16$ | | $1''.8 \pm 0''.22$ |
| | Modern observations | | | $1''.07 \pm 0''.06$ $1''.14 \pm 0''.11$ | |

| | | | | $1''.23 \pm 0''.04$ | |

In brackets presented estimates of the "true" lunar acceleration without distortion by apparent acceleration due to deceleration of the Earth rotation.

At first the comparison of de Sitter' estimations with the tidal theory brought disappointment to astronomers. Really, these estimations lead to the relation: $\dot{\sigma}/\dot{\sigma}_0 = 2.9 \pm 0.4$ that corresponds to $\alpha = N/(N + N_0) = 1.27 \pm 0.05$ in (15). But $\alpha > 1$ is senseless, corresponding to absurd $N > (N + N_0)$. Geophysical observations estimate in average $\alpha = 0.8$, that corresponds to $N/N_0 = 3.5 \div 4.7$. In its turn for $\alpha = 0.8$ Eq. 15 gives $\dot{\sigma}/\dot{\sigma}_0 = 6.8$ and with de Sitter' estimation $\dot{\sigma} = -5''.22 \pm 0''.30$ c$^{-2}$ it must be: $\dot{\sigma}_0 = 0''.77 \pm 0''.05$ c$^{-2}$. Therefore, astronomers concluded that in the equation of the Earth motion (12) an additional term of the type $\dot{\sigma}^* t_c^2$ must be included:

$$\dot{\sigma}^* t_c^2 = [\dot{\sigma}_0 - (0''.77 \pm 0''.05)] t_c^2 = +(0''.41 \pm 0''.09) t_c^2 \qquad (16)$$

However, the astronomical meaning of this correction that follows from the tidal theory is still a mystery for astronomers. The discovered divergence between the theoretical and observed accelerations of the Sun caused new thorough investigation of antique and modern estimates of solar and lunar accelerations. It turned out that these estimations also differ, and the difference between estimations of the Sun acceleration based on antique and modern observations approximately corresponds to (16):

$$\Delta \dot{\sigma}_0 = \dot{\sigma}_{0a} - \dot{\sigma}_{0c} = +(0''.55 \pm 0''.16) \text{ c}^{-2} \qquad (17)$$

Hereinafter the "a" and "c" respectively indicate characteristics estimated from historical ("antique") and modern observational data. To compare the lunar acceleration estimated on the basis of antique and modern data, "true" lunar accelerations not distorted by "apparent" acceleration due to non-uniform rotation of the Earth were used. In accordance with Eq. 15 true lunar acceleration is: $\dot{\sigma}_f = -\dot{\sigma} - \dot{\sigma}_0 n/n_0$. For example, to estimations of modern accelerations in Table 2 corresponds present true lunar acceleration: $\dot{\sigma}_{fc} = 5''.22 - 1''.23 \cdot 13.37 = -11''.22$ c$^{-2}$. Average difference in evaluations of the true Moon accelerations based on modern and antique data is:

$$\Delta \dot{\sigma}_{fa} = \dot{\sigma}_{fc} - \dot{\sigma}_{fa} = +(5''.7 \pm 2''.1) \text{ c}^{-2} \qquad (18)$$

This difference caused a long discussion and the study of an unusual hypothesis about the excess of the tidal part of lunar acceleration in ancient times over the modern acceleration by nearly 50 %. Interpretation of this difference in accordance with the first equation in (13) gave some astronomers the ground to claim that in ancient times the tidal influence of the Moon considerably exceeded the modern one. Contrary to astronomers' expectations, investigation of the Earth rotation fail to give an explanation of the differences in evaluation of the Earth and Moon accelerations based on antique and modern data; in fact, it added the new problem.

The tidal theory gives an estimation of true angular velocity of the Earth rotation with deduction of the Sun and Moon tidal influence. To much surprise of geophysicists, it turned out that after deduction of all main tidal influences the Earth angular velocity demonstrates secular increase, and its relative increase $\delta \omega / \omega$ is [1]: during the last 2200 years about $10 \cdot 10^{-8}$, and from 1700 till 1950 about $4 \cdot 10^{-8}$. As in the case of the Sun and Moon accelerations, evaluations of the relative increase of the Earth rotation according to antique and modern data also turned out to be different. During 20 centuries of the Christian Era, relative increase in the Earth rotation appeared approximately:

$$\delta\omega/\omega = (10-4)\cdot 10^{-8} = 6\cdot 10^{-8} \tag{19}$$

Discovered secular increase of the Earth rotation now does not have a recognized explanation although several ideas, suggesting a secular decrease of the inertia momentum of our planet have been examined.

In such a way during the three centuries of studies, most of the observed accelerations of the Sun and Moon motion got a qualitative and quantitative explanation in the theory of solar and lunar tides. However, hitherto no explanation has been found for the following specifics of the motion in the Earth-Moon system:
1. The origin of the secular acceleration of the Earth that is necessary for the coincidence of the tidal theory estimations with observations (16): $\dot{\sigma}^* t_c^2 = +(0''.41 \pm 0''.09)t_c^2$.
2. Secular acceleration of the Earth rotation that has led during 20 centuries to the relative increase of the Earth angular velocity (19): $\delta\omega/\omega \simeq 6\cdot 10^{-8}$.
3. The divergence (17) between the Sun acceleration estimates based on antique and modern observational data with the interval of 20 centuries between these epochs: $\Delta\dot{\sigma}_0 = \dot{\sigma}_{0a} - \dot{\sigma}_{0c} = +(0''.55 \pm 0''.16)$ c$^{-2}$.
4. The divergence (18) between evaluations of the tidal part of the Moon acceleration based on antique and modern data with the interval of 20 centuries between these epochs: $\Delta\dot{\sigma}_{fa} = +(5''.7 \pm 2''.1)$ c$^{-2}$.

Explanation for all these specifics of the Earth-Moon system kinematics lies in the non-stationarity of the space-time [5, 12]. The difference between Newtonian and decelerating physical time corresponds to Eq. 2 with the theoretical value of the Hubble constant [6, 11]: $H = 9\hbar G/16c^2 r_e^3 = 1.970\cdot 10^{-18}$ s$^{-1}$ (61.6 km/s/Mpc). In planetary astronomy it is convenient to use Hubble constant with the dimension (century)$^{-1}$ after multiplication by the number of seconds in a century: $H_c = 1.97\cdot 10^{-18} \cdot 3.156\cdot 10^9 = 6.22\cdot 10^{-9}$ c$^{-1}$. In the studies of relatively slow planetary processes it is necessary to introduce a correction for the cosmological change of the time scale. Eq. 12 for the Sun longitude with the cosmological correction in accordance with Eq. 2 has the form: $L = L_0 + n_0\tau_c + \delta L = L_0 + n_0 t_c + n_0 H_c t_c^2/2 + \delta L$. The term with Hubble constant is the additional acceleration (16) that was introduced by astronomers to adjust the acceleration tidal theory with observations. For the Sun motion $n_0 = 1''.3\cdot 10^8$ c$^{-1}$ and this cosmological correction is: $(\sigma^*)_\tau t_c^2 = n_0 H_c t_c^2/2 = [1''.3\cdot 10^8 \cdot 6.22\cdot 10^{-9}/2]t_c^2 = +0''.404 t_c^2$, almost exactly coinciding with (16). Apparent acceleration of the Earth rotation also reflects cosmological growth of the time-scale: $\tau_c - t_c = H_c t_c^2/2$. This relation can be used to evaluate the relative growth of the Earth angular velocity during 20 centuries: $(\delta\omega/\omega)_\tau = (\tau_c - t_c)/t_c = H_c t_c/2 = 6.22\cdot 10^{-9}\cdot 20/2 = 6.22\cdot 10^{-8}$, that is also corresponds quite well to the estimation (19) based on astronomical observations.

In the case of relatively short time intervals, for example, for calculations based on astronomical observations for the last 250 years, cosmological corrections are less than mean observation errors. However, in calculations based on the antique data of 20 centuries before our time, such corrections already exceed the level of observation errors. This is why estimation of the Sun acceleration based on antique data ought to be done with correction for the shorter physical time-scale in the past. Prior to introduction of this cosmological correction one should first estimate the value of apparent relative acceleration of the Earth rotation: $(\dot{\omega}/\omega)_\tau = (\tau_c - t_c)/t_c^2 = H_c/2 = 3.11\cdot 10^{-9}$ c$^{-1}$. Then using Eq. 14 one can get: $(\dot{\sigma}_0)_\tau = -n_0(\dot{\omega}/\omega)_\tau = -n_0 H_c/2 = -1''.3\cdot 10^8 \cdot 3.11\cdot 10^{-9} = -0''.404$ c$^{-2}$. By adding this correction to the right part of (17) we immediately eliminate the divergence in the Sun acceleration estimates based on antique and modern observations.

It is possible to introduce the cosmological correction for the Moon acceleration by the same method using the known value of the average Moon motion: $n = 13.37 n_0 = 1''.74\cdot 10^9$ c$^{-1}$ and the cosmological correction for relative acceleration of the Earth rotation: $(\dot{\sigma})_\tau = -n(\dot{\omega}/\omega)_\tau = -nH_c/2 = -1''.74\cdot 10^9 \cdot 3.11\cdot 10^{-9} = -5''.41$ c$^{-2}$. By adding this cosmological correction to the right part of (18) we eliminate also the divergence in estimations of the Moon acceleration for antique and modern observations. All these cosmological corrections for the Moon and Earth motions presented in the Table 3.

*Table 3. Cosmological corrections in kinematics of the Earth-Moon system*

|  | Estimations based on observations | Cosmological corrections | $\delta(\%)$ |
|---|---|---|---|
| Additional orbital acceleration of the Earth in the tidal theory (16). (seconds of arc) (century)$^{-2}$ | $0''.41 \pm 0''.09$ | $(\sigma^*)_\tau = n_0 H_c / 2 = 0''.404$ | + 1.5 |
| Divergence in estimations of the Sun acceleration based on antique and modern data (17). (seconds of arc) (century)$^{-2}$ | $-0''.55 \pm 0''.16$ | $(\dot\sigma_0)_\tau = -n_0 H_c / 2 = -0''.404$ | - 26 |
| Divergence in estimations of the Moon acceleration based on antique and modern data (18). (seconds of arc) (century)$^{-2}$ | $-5''.7 \pm 2''.1$ | $(\dot\sigma)_\tau = -n H_c / 2 = -5''.41$ | + 5.1 |
| Apparent relative increase of the Earth angular velocity during 20 centuries (19) | $6 \cdot 10^{-8}$ | $(\delta\omega/\omega)_\tau = H_c t_c / 2 = 6.22 \cdot 10^{-8}$ | + 3.7 |

$\delta(\%)$ is the deviation of the theoretical corrections from observations (on average, in %)

Despite the deviations about $1.5 \div 26\%$ (on average 9%) of theoretical cosmological corrections from estimates based on observations, the coincidence of the theory under discussion with observations can be considered quite satisfactory because standard deviations of astronomical data, concerning accelerations in the Earth-Moon system, are 30%, on average.

It is interesting to solve the inverse problem. Using experimental estimations (16 – 19) and cosmological corrections in the Table 3 one can evaluate Hubble constant from the accelerations in the Earth-Moon system: $H_c = (6.83 \pm 1.11) \cdot 10^{-9}$ c$^{-1}$, that corresponds to $H = (2.16 \pm 0.35) \cdot 10^{-18}$ s$^{-1}$ = $(67.5 \pm 11)$ km/s/Mpc.

Physical nature of introduced cosmological corrections in the kinematics of the Earth-Moon system is the same as in the phenomenon of the apparent acceleration of the universe expansion: uniform and moderately decelerating motions in the physical time appear accelerated, if to use for estimations invariable Newtonian time with uniform scale.

### 3. Isotope decay kinetics in non-stationary universe

Non-stationary state of the space-time determining cosmological deceleration of the course of time can be observed not only on the cosmic scale in astrophysical studies but also in microcosm. The cosmological deceleration of physical time in relation to the invariable and uniform scale of Newtonian time becomes obvious in the analysis of decay kinetics of radioactive isotopes [5, 9, 10].

The isotope decay constant is proportional to the probability of the isotope unstable nucleus decomposition, which, in its turn, is inversely proportional to the mean lifetime of the isotope nuclei activated state. In accordance with the concept of physical time, the mean lifetime of isotope nuclei must be estimated not in the Newtonian but in the physical time. If isotopes actually decay in the decelerating physical time, the difference between the actual quantity of the decayed nuclei and a prediction made with kinetics equation with Newtonian time will be growing with the increase of the isotope half-decay time. To derive the law of the radioactive decay in the physical time, one needs to replace the decay constant for Newtonian time $\lambda_t^I$ by the decay constant for physical time $\lambda_\tau^I$. The values of constants $\lambda_t^I$ and $\lambda_\tau^I$ are not equal, and $\lambda_t^I > \lambda_\tau^I$. Kinetics of the isotope concentration changes in physical time is defined by the modified law of radioactive decay:

$$N_\tau^I = N_0^I \exp\left(-\lambda_\tau^I \tau\right) = N_0^I \exp[-\lambda_\tau^I (t + Ht^2/2)] \qquad (20)$$

Here $N_0^I$, $N_\tau^I$ are concentrations of the isotope I at the initial and current time respectively, $\lambda_\tau^I$ is the decay constant of isotope I in physical time. Because of relatively short periods (not more than 20 years) of experimental determination of decay constants of long-living isotopes, it is impossible to take into account the deceleration of physical time. Therefore the decay constants used in calculations are, in fact, overestimated values of the decay constants for decelerating physical time. The difference between calculations in the Newtonian and physical time becomes significant only for the long time intervals, for example, in isotope geochronology. In addition to the law of radioactive decay, the methodology of isotope geochronology uses the relation:

$$t_e^I = L_e^I / \lambda_t^I \qquad (21)$$

Here $\lambda_t^I$ is the decay constant of the isotope I specifying the method of isotope geochronology, $t_e^I$ is experimental estimate of the Newtonian age of a geological sample, and $L_e^I$ is a kinetics function of experimentally estimated isotope concentrations formed in the process of the main isotope I decay. Please, note that the index "e" denotes experimental estimations of the characteristics. The type of function $L_e^I$ depends on the specifics of a concrete method of isotope geochronology. Isotope geochronology employs many modifications of such formulae. However, the important thing is that because of the permanent use of the radioactive decay law, the formulae like Eq. 21 used for the age evaluation always include the main isotope decay constant $\lambda_t^I$ as *the denominator*.

Comparison of the formulae (20, 21) for Newtonian and physical times leads to the relation:

$$\tau_e^I - t_e^I = \tau_e^I / \lambda_t^I (\lambda_t^I - \lambda_\tau^I) \qquad (22)$$

One may see from this relation that the difference between age evaluations in the physical and Newtonian times increases with the age of the sample and with using of a more long-living isotope (with lesser decay constant) for the analysis. If to analyze one and the same geological sample having only one actual age in the physical time using various isotope methods, estimations of its Newtonian age will not coincide. Moreover, the method employing a more long-living isotope will always give a lower value of the sample Newtonian age.

It makes sense to perform study of the isotope age divergences caused by the difference between Newtonian and physical time scales in methods employing isotopes with big difference between decay constant values. Among such methods there are widely used in isotope geochronology (U-Pb) (the decay constant of $^{238}$U: $\lambda_t^{238} = 1.55 \cdot 10^{-10}$ year$^{-1}$) and (Rb-Sr) (the decay constant of $^{87}$Rb: $\lambda_t^{87} = 1.42 \cdot 10^{-11}$ year$^{-1}$) methods. From thousands published in recent years isotope age estimations one can choose tenths publications reporting results of the parallel use (U-Pb) and (Rb-Sr) methods. The divergence in sample age estimations by these two methods is especially contrast in samples over a billion years of age. Study of the sample Newtonian ages estimated with parallel use of (U-Pb) and (Rb-Sr) methods demonstrates their systematic divergence with the mean difference around 0.17 billion years (+ 6.5 %). This divergence in age evaluations significantly exceeds the mean analytical error level in isotope geochronology (1-2 %). Evaluations of the sample Newtonian ages $t_{ei}^{238}$ and $t_{ei}^{87}$ allow to estimate experimental values of the functions $L_{ei}^{238}$ и $L_{ei}^{87}$ using Eq. 21. Then, using again Eq. 21 it is possible to estimate experimental values of the decay constants in physical time $\lambda_{\tau e}^{238}$ and $\lambda_{\tau e}^{87}$, providing a minimal sum of squares of the geological sample age differences in physical time: $\Phi = \sum_{(i)} \left( \frac{L_{ei}^{238}}{\lambda_{\tau e}^{238}} - \frac{L_{ei}^{87}}{\lambda_{\tau e}^{87}} \right)^2$. For the analyzed data (see e.g. Table A6 in [6]) estimations of the decay constants, evaluated by the gradient descend minimization method are the following: $\lambda_{\tau e}^{238} = 1.458 \cdot 10^{-10}$ year$^{-1}$; $\lambda_{\tau e}^{87} = 1.25 \cdot 10^{-11}$ year$^{-1}$.

Fig. 2 presents the differences in Newtonian age estimates of geological samples of the Earth and Moon rocks determined by (Rb-Sr) (filled circles near the line 1) and (U-Pb) (filled squares near the line 2) isotope methods.

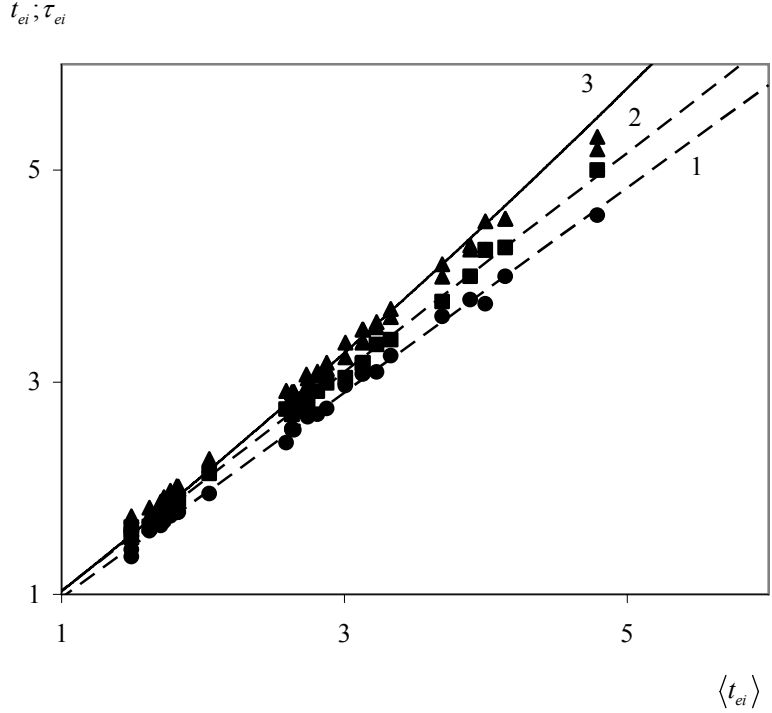

*Fig. 2. The estimations of the geological sample ages by (U-Pb) and (Rb-Sr) isotope methods.*

Abscissas in Fig. 2 represent estimations of the average sample Newtonian age: $\langle t_{ei} \rangle = (t_{ei}^{238} + t_{ei}^{87})/2$ in billion years. Ordinates represent the sample age estimates in billion years in Newtonian $(t_{ei})$ and physical time $(\tau_{ei})$. Estimations of sample physical age evaluated by Eq. 21 with decay constants in physical time are plotted in Fig. 2 by filled triangles near the curve 3. Analysis shows that the mean divergence in the sample physical ages evaluated in parallel using of (U-Pb) and (Rb-Sr) methods is less than 0.01 billion years (– 0.2 %) that is lower by more than an order as compared to the mean difference in sample Newtonian ages.

The concept of the physical time deceleration gives a unique opportunity to estimate the value of the Hubble constant using isotopic ages of the Earth and Moon rocks instead of the redshifts in spectra of distant galaxies. To estimate Hubble constant Eq. 2 can be used in the form: $\tau_{ei} = t_{ei} + H_e t_{ei}^2/2$ and the value of the constant is calculated from the condition: $S = \sum_{(i)} (\tau_{ei} - t_{ei} - H_e t_{ei}^2/2) = 0$. For the analyzed data the estimation of Hubble constant is: $H_e = (65.4 \pm 6)$ km/s/Mpc $(2.1 \pm 0.19) \cdot 10^{-18}$ s$^{-1}$; 0.066 Gyr$^{-1}$). This estimation is close to the theoretical value and corresponds to adopted in astrophysics the Hubble constant range: $50 \div 80$ km/s/Mpc. In Fig. 2 the relative standard deviations of isotopic estimations of the geological sample physical age from Eq. 2 (curve 3 in Fig. 2) is under 5 %. The main cause of observed divergence in the age estimations of geological samples from the Earth and the Moon received by means of using isotopes with different decay constants is the difference between Newtonian and physical time scales in the non-stationary universe.

Coincidence of the Hubble constant estimations based on the isotope geochronology data and kinematics parameters of the Earth-Moon system with estimates based on the extragalactic object observations is an expressive confirmation of a phenomenon of cosmological deceleration of the course of physical time.